\documentclass[prl,aps,preprintnumbers,nofootinbib,10pt,twocolumn]{revtex4}
\usepackage{stmaryrd}
\usepackage{epsfig}
\usepackage{amsfonts}
\usepackage{amsmath}
\usepackage{slashed}
\usepackage{graphicx}
\usepackage{color}
\usepackage{amssymb,amsmath,psfrag,slashed,graphicx}
\usepackage{caption}
\usepackage{subfigure}

\allowdisplaybreaks[4]
\newif\ifContLineOne
\newif\ifContLineTwo
\newif\ifContLineThree

\begin{document}

\preprint{JLAB-THY-19-3025}

\title{$B$-meson light-cone distribution amplitude from the Euclidean quantity}

\author{Wei Wang$^1$~\footnote{correspondence author, wei.wang@sjtu.edu.cn},
Yu-Ming Wang$^2$~\footnote{correspondence author, wangyuming@nankai.edu.cn},
Ji Xu$^{1,3}$~\footnote{correspondence author, xuji1991@sjtu.edu.cn},
and Shuai Zhao$^{4,5}$~\footnote{correspondence author, szhao@odu.edu}}
\affiliation{$^1$ INPAC, SKLPPC, MOE KLPPC,
School of Physics and Astronomy, Shanghai Jiao Tong University, Shanghai 200240,   China \\
$^2$ School of Physics, Nankai University, 300071 Tianjin, China\\
$^3$ Nuclear Science Division, Lawrence Berkeley National Laboratory, Berkeley, CA 94720, USA \\
$^4$ Physics Department, Old Dominion University, Norfolk, VA 23529, USA \\
$^5$ Theory Center, Thomas Jefferson National Accelerator Facility, Newport News, VA 23606, USA
}

\date{\today}


\begin{abstract}

A new method for the model-independent determination of the light-cone distribution amplitude (LCDA) of the
$B$-meson  in heavy quark effective theory (HQET) is proposed by combining the large momentum
effective theory (LaMET) and the numerical simulation technique on the  Euclidean lattice.
We demonstrate the autonomous scale dependence of the non-local quasi-HQET operator with the aid of the
auxiliary field approach, and further determine the perturbative matching coefficient entering the hard-collinear
factorization formula for the $B$-meson quasi-distribution amplitude at the one-loop accuracy.
These results will be crucial to explore the partonic structure of heavy-quark hadrons in the static limit
and to improve the theory description of exclusive $B$-meson decay amplitudes based upon perturbative
QCD factorization theorems.

\end{abstract}

\maketitle

{\bf Introduction:} $B$-meson light-cone distribution amplitude (LCDA) in
heavy-quark effective theory (HQET)\cite{Grozin:1996pq}
serves  as an indispensable ingredient  for establishing  QCD factorization theorems of
exclusive $B$-meson decay amplitudes \cite{Beneke:1999br,Beneke:2000wa,Beneke:2001at,Bosch:2001gv,Becher:2005fg}
and for constructing  light-cone sum rules of numerous hadronic matrix elements,
whose factorization properties are not yet completely explored
at leading power in the heavy-quark expansion,  from the vacuum-to-$B$-meson
correlation functions \cite{Khodjamirian:2006st,DeFazio:2005dx,Wang:2015vgv,Wang:2017jow,Lu:2018cfc,Gao:2019lta}.
Defined as the light-ray matrix elements of the composite HQET quark-gluon  operators,
they encode information of the non-perturbative strong interaction dynamics from the soft-scale fluctuation
of the $B$-meson system and our limited knowledge of these distribution amplitudes has
become the major stumbling block for precision calculations of the $B$-meson  decay observables,
which are crucial to disentangle the Standard-Model (SM) background contributions from
the  genuine new physics effects.

Model-independent properties of the leading-twist $B$-meson LCDA $\phi_B^{+}(\omega, \mu)$,
including its renormalization group equation (RGE) and  perturbative QCD constraints at large $\omega$,
have received increasing attention in  recent years
\cite{Lange:2003ff,Bell:2013tfa,Braun:2014owa,Braun:2019wyx,Lee:2005gza,Kawamura:2008vq,Feldmann:2014ika}.
By contrast, nonperturbative determinations of $\phi_B^{+}(\omega, \mu)$
have been mainly performed in the framework of the QCD sum rules (QCDSR) invoking the
quark-hadron duality ansatz \cite{Braun:2003wx}, where both the perturbative
corrections to the leading-power contribution and the subleading-power contributions
from  quark-gluon condensate operators were taken into account systematically.
One main drawback of constructing the phenomenological models
for the $B$-meson distribution amplitude $\phi_B^{+}(\omega, \mu)$ from the classical QCDSR method  lies in the fact that
the light-cone separation between the  effective heavy-quark field and the light anti-quark field
needs to be sufficiently small (of order $1 \sim 3 \, {\rm GeV}^{-1}$) to
guarantee the validity of the local operator-product-expansion (OPE)  for the HQET  correlation function under discussion.
In addition, meaningful constraints on the first inverse moment
$\lambda_B^{-1}(\mu)$ \cite{Beneke:2011nf,Wang:2016qii,Wang:2018wfj,Beneke:2018wjp}
can be obtained by measuring the integrated branching fractions of the radiative leptonic
$B$-meson decays with a photon energy cut $E_{\gamma} \geq E_{\rm cut}$ at the Belle II experiment.
Keeping in mind that the exact RGE for the inverse moment $\lambda_B^{-1}(\mu)$ involves all the logarithmic
moments $\sigma_B^{(n)}(\mu)$ in perturbation theory, the precise shape of the $B$-meson distribution amplitude,
in particular its small-momentum behaviour, cannot be controlled by
a single non-perturbative parameter $\lambda_B^{-1}(\mu)$ to a good approximation.
 It is  then evident that determining the momentum dependence of the $B$-meson LCDA $\phi_B^{+}(\omega, \mu)$
with model-independent techniques  is of top priority for the precision descriptions of  exclusive $B$-meson decays.

Performing the lattice QCD calculation of the leading-twist distribution amplitude $\phi_B^{+}(\omega, \mu)$
directly is known to be complicated by the appearance of the light-cone separated quark fields defining
the very HQET matrix element for a long time.
A promising approach to circumvent this long-standing problem has been recently proposed under the name of the
large momentum effective theory (LaMET) by X. Ji \cite{Ji:2013dva,Ji:2014gla}
(see also \cite{Cichy:2018mum,Zhao:2018fyu,Ji:2020ect} for a review).
The essential strategy of this novel proposal  consists in the construction of a time-independent quasi-quantity
which, on the one hand, can be readily computed  on a Euclidean lattice and, on the other hand, approaches
the original hadronic distribution amplitude on the light cone under Lorentz boost.
In this Letter, we implement Ji's proposal to extract the leading-twist LCDA $\phi_B^{+}(\omega, \mu)$ of the $B$-meson
in HQET  by demonstrating the multiplicative renormalization of the constructed quasi-HQET operator to all orders in perturbation theory,
by determining the short-distance function appearing in the hard-collinear factorization formula of the $B$-meson
quasi-distribution amplitude, and by exploring  future opportunities  of lattice QCD calculations.

{\bf $B$-meson (quasi)-distribution amplitudes:}
The leading-twist LCDA $\tilde{\phi}_B^{+}(\eta, \mu)$ in coordinate space is defined by
the renormalized HQET matrix element of a  light-ray soft  operator \cite{Lange:2003ff}
\begin{eqnarray}
&& \langle  0 | \left (\bar q_s \, Y_s \right)(\eta \, \bar{n}) \,  \slashed {\bar n} \, \gamma_5 \,
\left (Y_s^{\dag} \, h_v \right )(0)| \bar B(v)\rangle \nonumber \\
&& = i \tilde f_B(\mu) \, m_B \, \tilde{\phi}_{B}^{+}(\eta, \mu) \,,
\label{def: twist-2 B-meson DA}
\end{eqnarray}
where the soft light-cone ($\bar n^2=0$) Wilson line is given by
\begin{eqnarray}
Y_s(\eta \, \bar n)= {\rm P} \, \left \{ {\rm  Exp} \left [   i \, g_s \,
\int_{- \infty}^{\eta} \, dx \,  \bar n  \cdot A_{s}(x \, \bar n) \right ]  \right \} \,,
\label{def: soft Wilson line}
\end{eqnarray}
and the static decay constant $\tilde f_B(\mu)$ of the $B$-meson can be expressed in terms of
$f_B$ in QCD \cite{Beneke:2005gs}.
Applying the Fourier transformation for  $\tilde{\phi}_{B}^{+}(\eta, \mu)$ leads to
the momentum-space distribution function \cite{Braun:2003wx}
\begin{eqnarray}
\phi_{B}^{+}(\omega, \mu) = {1 \over 2 \, \pi} \, \int_{- \infty}^{+ \infty} \, d \eta \,
e^{i \, \, \bar n \cdot v  \, \omega \,  \eta} \, \tilde{\phi}_{B}^{+}(\eta - i \, \epsilon, \mu) \,.
\end{eqnarray}
Following the construction presented in  \cite{Ji:2013dva},
we will employ the following $B$-meson quasi-distribution amplitude
(see also \cite{Kawamura:2018gqz})
\begin{eqnarray}
&&i \tilde f_B(\mu) \, m_B \, \varphi_{B}^{+}(\xi, \mu)
=  \int_{- \infty}^{+ \infty} \, { d \tau \over 2 \, \pi }\,
e^{i \, n_z \cdot v  \, \xi \,  \tau} \, \nonumber \\
&&  \langle  0 | \left (\bar q_s \, Y_s \right)(\tau \, n_z) \,  \not \! n_z \, \gamma_5 \,
\left (Y_s^{\dag} \, h_v \right )(0)| \bar B(v)\rangle \,,
\label{def: twist-2 quasi-B-meson DA}
\end{eqnarray}
defined by the spatial correlation function of two collinear (effective)-quark fields
with $n_z=(0, 0, 0, 1)$.
For the sake of demonstrating  QCD factorization for the quasi-distribution amplitude
$\varphi_{B}^{+}(\xi, \mu)$,  we will work in a Lorentz boosted frame of the $B$-meson
with $\bar n \cdot v \gg n \cdot v$ and set $v_{\perp \mu}=0$ without loss of generality.
As a consequence, only the ultra-collinear gluons  couple with the boosted heavy quark
in the low-energy effective theory and the soft Wilson lines $Y_s(\theta \, u)$ will be substituted by
the ultra-collinear Wilson lines $W_n(\theta \, u)$ in the boosted HQET (bHQET) accordingly  \cite{Fleming:2007qr}.

{\bf Multiplicative renormalization:}
To facilitate the lattice QCD evaluation of the quasi-distribution amplitude $\varphi_{B}^{+}(\xi, \mu)$,
it is of vita importance to show that such quasi-quantity will renormalize  multiplicatively  to all orders
in perturbation theory applying the lattice regularization scheme.
For this purpose, it has proven to be most convenient employing the one-dimensional auxiliary field formalism
for the contour integrals introduced in \cite{Gervais:1979fv}.
The resulting Lagrangian for the  ultra-collinear gluon interactions with both the effective bottom-quark field $h_v$ and
the auxiliary field ${\cal Q}$ can be written as
\begin{eqnarray}
\cal{L}=& {\cal L}_{\rm{bHQET}}+\bar{{\cal Q}}(x) \, (i n_z \cdot D_n - \delta m ) \, {\cal Q}(x),
\end{eqnarray}
where the ``dynamical" mass term originates from the self-energy correction to the $ {\cal Q}$ field
in the dimensionful cut-off scheme \cite{Dotsenko:1979wb}, in analogy  to the scheme-dependent residual mass term
in the HQET formalism \cite{Falk:1992fm,Maiani:1991az,Beneke:1994sw},
and the ultra-collinear covariant derivative $D_{n}^{\mu}= \partial^{\mu} - i \, g_s \, T^a \, A_{n}^{a, \, \mu}$.
Alternatively,  the ultraviolet (UV) linear divergences from the Wilson-line corrections
in (\ref{def: twist-2 quasi-B-meson DA}) can be removed by introducing  the proper subtraction term
defined by a simpler matrix element but with the same power divergences \cite{Radyushkin:2017cyf,Orginos:2017kos,Braun:2018brg}.
It is straightforward to rewrite the non-local operator defining the $B$-meson quasi-distribution amplitude as
follows \cite{Craigie:1980qs,Ji:2017oey}
\begin{eqnarray}
{\cal O}(\tau \, n_z, 0) &=& \left [ \bar \chi_n(\tau \, n_z) \,  \not \! n_z \, \gamma_5 \,  {\cal Q}(\tau \, n_z) \right ]
\, \left [\bar {\cal Q}(0) \, h_v(0) \right ]  \nonumber \\
&\equiv&  {\cal J}_{\chi \cal Q}(\tau \, n_z) \,\,  {\cal J}_{{\cal Q}  h_v}(0) \,,
\end{eqnarray}
where $\chi_n$ stands for the ultra-collinear quark field.

Thanks to the heavy-quark spin symmetry and the light-quark chiral symmetry
for the effective Lagrangian $\cal{L}$,
both of the two currents ${\cal J}_{\chi \cal Q}$ and ${\cal J}_{{\cal Q}  h_v}$
renormalize multiplicatively under radiative corrections  \cite{Manohar:2000dt}
\begin{eqnarray}
{\cal J}_{\chi \cal Q}(\tau \, n_z) &=& Z_{\chi \cal Q}^{(R)} \, {\cal J}^{(R)}_{\chi \cal Q}(\tau \, n_z, \mu) \,, \nonumber \\
\qquad  {\cal J}_{{\cal Q}  h_v}(0)  &=&  Z_{{\cal Q}  h_v}^{(R)} \, {\cal J}^{(R)}_{{\cal Q}  h_v}(0, \mu)   \,,
\end{eqnarray}
at all orders in $\alpha_s$.
We are therefore led to conclude the autonomous renormalization of the composite non-local operator
${\cal O}(\tau \, n_z, 0)$, namely
\begin{eqnarray}
{\cal O}(\tau \, n_z, 0) = Z_{\chi \cal Q}^{(R)} \, Z_{{\cal Q}  h_v}^{(R)}  \,  {\cal O}^{(R)}(\tau \, n_z, 0, \mu).
\end{eqnarray}
It needs to be stressed that such multiplicative-renormalization property holds in both dimensional regularization
and lattice regularization schemes due to the reparametrization invariance of the heavy quark mass \cite{Falk:1992fm},
which can be readily understood by introducing the generalized covariant derivative
$i {\cal D}^{\mu} = i D^{\mu} + \delta m \, n_{z}^{\mu}$.
In general, the renormalized non-local quasi-operator ${\cal O}^{(R)}$ for a given regularization scheme violating
translation  invariance (including but not limited to the lattice regularization scheme)
can be expressed as \cite{Ji:2017oey,Zhang:2018diq}
\begin{eqnarray}
 {\cal O}^{(R)}(\tau \, n_z, 0, \mu) = \left [  Z_{\chi \cal Q}^{(R)} \, Z_{{\cal Q}  h_v}^{(R)}  \right ]^{-1} \,
 e^{\overline{\delta m} \, \tau} \,  {\cal O}(\tau \, n_z, 0),
\end{eqnarray}
with the imaginary mass $\delta m= i \, \overline{\delta m}$ due to the space-like gauge vector $n_z$ \cite{Polyakov:1980ca}.

\begin{figure}[http]
\includegraphics[width=3.2 in]{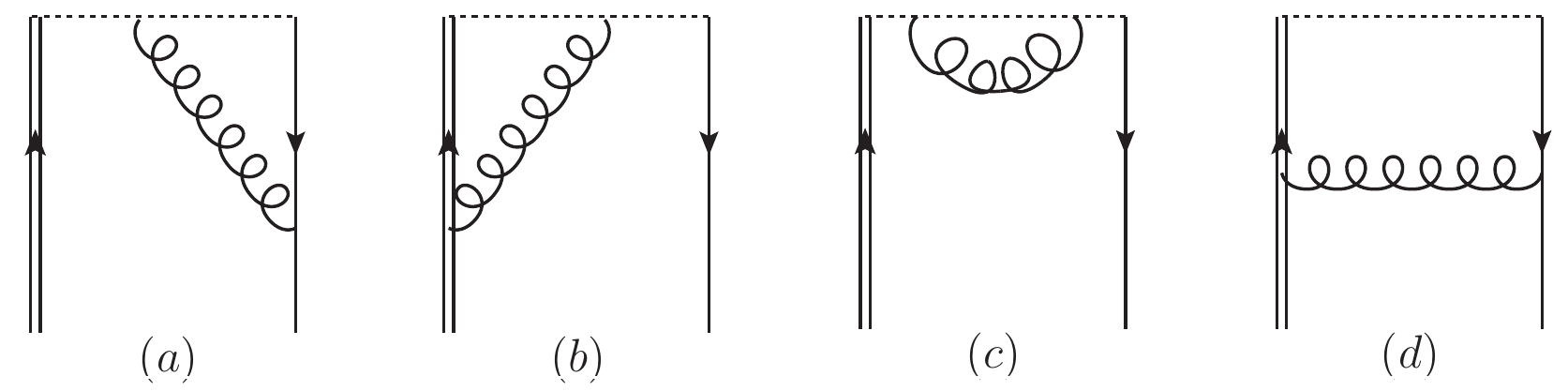}
\caption{One-loop corrections to the quasi-distribution amplitude of the $B$-meson
 $\varphi_B^{+}(\xi, \mu) $: the effective HQET  bottom quark is represented by the double line,
 and the spacelike Wilson line is indicated by the dashed line.}
\label{1-loop-matching}
\end{figure}

{\bf Hard-collinear factorization formula:}
We now proceed to determine the perturbative matching coefficient function entering the
hard-collinear factorization formula for $ {\cal O}^{(R)}(\tau \, n_z, 0)$ at $\tau \ll 1/\Lambda_{\rm QCD}$
\begin{eqnarray}
{\cal O}^{(R)}(\tau \, n_z, 0, \mu) &=& \int d \eta \, \tilde{H}(\tau, \, \eta, \, n_z \cdot v, \, \mu) \,
 P^{(R)}(\eta \, \bar n, 0, \mu), \nonumber \\
 P^{(R)}(\eta \, \bar n, 0, \mu) &=&  \left [ \left ( \bar \chi_n \, W_n \right)(\eta \, \bar  n) \,  \slashed {\bar n} \, \gamma_5 \,
\left (W_n^{\dag} \, h_v \right ) (0)  \right  ]^{(R)},  \nonumber
\end{eqnarray}
at leading power in $\Lambda_{\rm QCD} \, \tau$, which can be  Fourier-transformed into the momentum-space relation
\begin{eqnarray}
\varphi_B^{+}(\xi, \mu) &=& \int_0^{\infty} d \omega \, H(\xi, \, \omega, \, n_z \cdot v, \, \mu)
\, \phi_{B}^{+}(\omega, \mu) \nonumber \\
&& + \, O \left ({\Lambda_{\rm QCD} \over n_z \cdot v \, \xi} \right )  \,.
\label{QCDF-formula}
\end{eqnarray}
Applying the default power counting scheme one can readily identify that the hard correction from the
1-loop box diagram in figure \ref{1-loop-matching} is power suppressed and it will therefore give rise to the vanishing contribution
to the  perturbative matching function  $H$.
We further verify explicitly that the collinear contribution to the quasi-distribution amplitude is precisely
reproduced by the corresponding diagrams for the $B$-meson LCDA at one loop.
The obtained hard function at $O(\alpha_s)$ reads
\begin{widetext}
\begin{eqnarray}
H(\xi, \, \omega, \, n_z \cdot v, \, \mu) &=& \delta (\xi-\omega)
+ {\alpha_s(\mu) \, C_F \over 4 \, \pi} \,
\bigg \{ {1 \over \omega - \xi} \, \left [ 3 - 2 \, \ln \left ( {\mu \over 2 \, n_z \cdot v \, (\omega-\xi)} \right )
- {2 \,  \xi \over \omega}  \, \ln \left ( {\xi \over \xi - \omega} \right ) \right ] \, \theta(-\xi) \,\, \theta(\omega) \nonumber \\
&& + \, \left \{ {1 \over \omega - \xi} \, \left [   3 - 2 \, \left (1 +  {2 \, \xi \over \omega}  \right ) \,
\ln \left ( {\mu \over 2 \, n_z \cdot v \, (\omega-\xi)} \right )
- {2 \,  \xi \over \omega}  \, \left (\ln \left ( {\omega - \xi \over \xi} \right )  + 1 \right ) \right ] \right \}_{\oplus}
\, \theta(\xi) \,\, \theta(\omega- \xi)    \nonumber \\
&& + \, \left \{  \, {1 \over  \xi -\omega} \,  \left [ 3 - 2 \, \ln \left ( {\mu \over 2 \, n_z \cdot v \, (\xi-\omega)} \right )
-  {2 \,  \xi \over \omega}  \,  \ln \left ( {\xi \over \xi - \omega} \right )  \right ] \right \}_{\oplus}
\, \theta(\omega) \,\, \theta(\xi-\omega)  \nonumber \\
&& +  \, 2 \,  \left [ \ln^2 {\mu \over n_z \cdot v \,\, \xi} - 3 \, \ln {\mu \over n_z \cdot v \,\, \xi} + f(a)  \right ]  \, \delta(\xi-\omega)  \bigg \} \,,
\end{eqnarray}
\end{widetext}
where the plus distribution is defined by (with $a > 1$)
\begin{eqnarray}
\left \{ {\cal F}(\xi, \omega) \right \}_{\oplus} =  {\cal F}(\xi, \omega)
- \delta(\xi - \omega) \, \int_0^{a \, \xi}  \, d t \,  {\cal F}(\xi, t) \,,
\end{eqnarray}
and the subtraction-scheme dependent term
\begin{eqnarray}
f(a) &=&  \ln\frac{a^2}{4(a-1)^3} \, \ln \frac{\mu}{n_z\cdot v \,\, \xi}
+ \ln(a-1) \, \ln  \frac{8 \, (a-1)}{a} \nonumber \\
&& + \,  {\rm Li}_2(1-a) + \ln a \, \ln \left (\frac{a}{4} \right ) - {1 \over 2} \ln(a-1)   \nonumber \\
&& + \, \ln (8 \, a) + \ln^2 2 + \frac{\pi^2}{8}  - 3
\end{eqnarray}
will compensate the same scheme dependence of the newly introduced  plus distribution
for the convolution of the hard function $H$ with a smooth test function.
An advantage of introducing the above-mentioned plus function is that
it allows to implement both the ultraviolet and infrared subtractions for the perturbative
matching procedure simultaneously.
Distinguishing the ultraviolet renormalization scale $\nu$ of the composite quasi-operator ${\cal O}^{(R)}$
from the factorization scale $\mu$ of separating the hard and collinear strong interaction dynamics
for this quantity,
it is straightforward to demonstrate a complete cancellation of the $\mu$-dependence for the factorization
formula of $\varphi_B^{+}(\xi, \nu, \mu)$ at  one loop, by employing the Lange-Neubert evolution equation of
the $B$-meson distribution amplitude $\phi_{B}^{+}(\omega, \mu)$ \cite{Lange:2003ff}.

\begin{figure}[htbp]
\centering
\includegraphics[width=0.80\columnwidth]{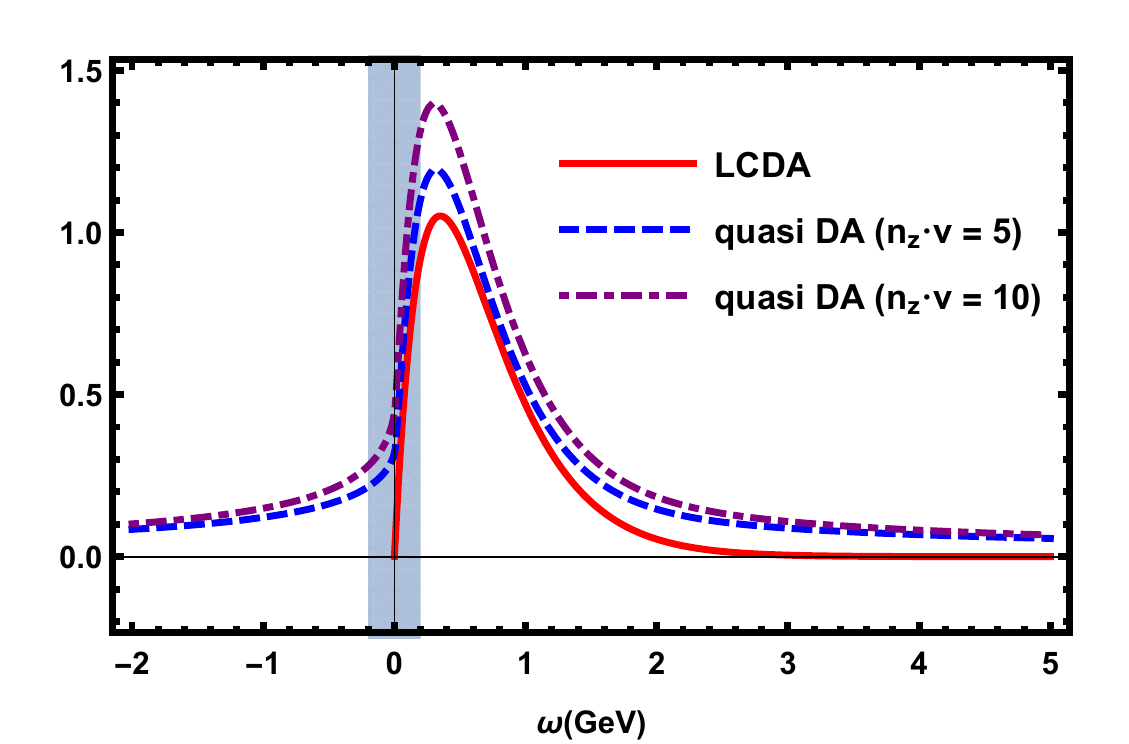}
\includegraphics[width=0.80\columnwidth]{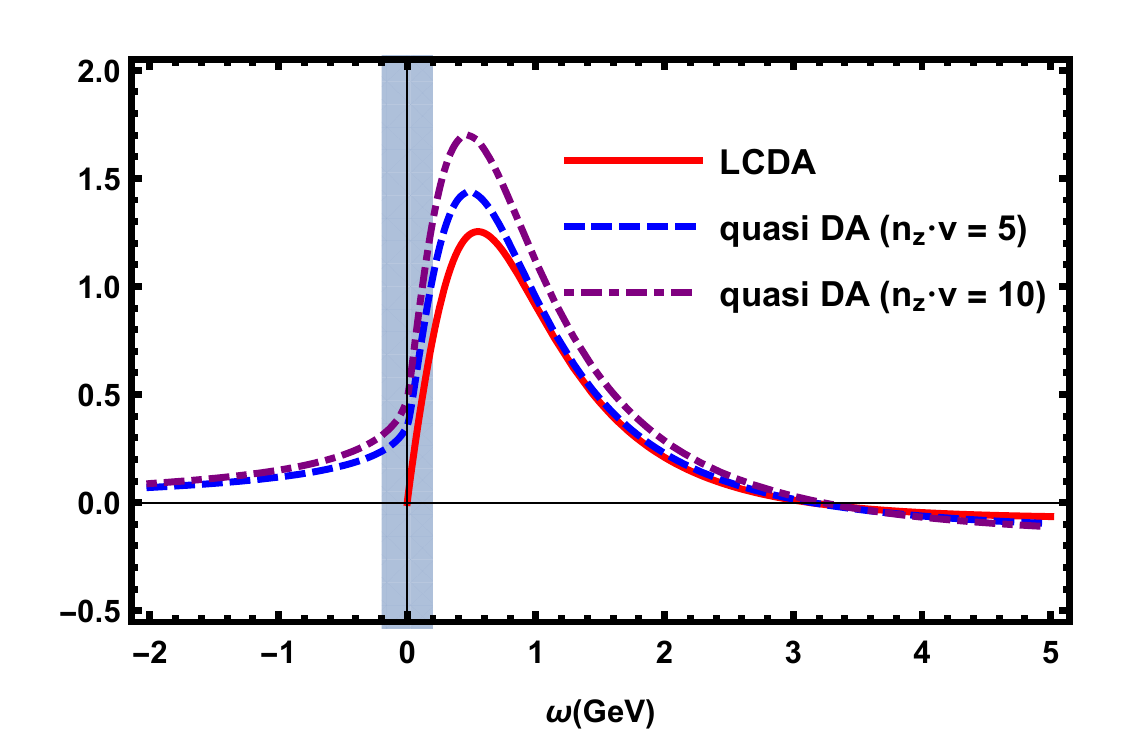}
\centering
\caption{The resulting $\omega$-shapes of the $B$-meson quasi-distribution amplitude
$\varphi_B^{+}(\xi = \omega, \mu=1.5 \, {\rm GeV})$
in bHQET from the hard-collinear factorization theorem (\ref{QCDF-formula}) and from the two non-perturbative models
of  $\phi_{B}^{+}(\omega, \mu=1.5 \, {\rm GeV})$ presented in (\ref{two models of B-meson LCDA}),
with two different values of $n_z \cdot v$.
The shadow region of $|\omega| \leq 200 \,  {\rm MeV}$ is excluded due to
inapplicability of the hard-collinear factorization formula (\ref{QCDF-formula})
for $|n_z \cdot v \,\,  \omega| \leq 1.0 \,  {\rm GeV}$.}
\label{models of quasi-DA}
\end{figure}

{\bf Perspectives for lattice calculations:}
An important step in obtaining the $B$-meson LCDA in bHQET based upon
Ji's approach is to perform the lattice QCD simulation for the spatial correlation
$\varphi_B^{+}(\xi, \mu)$ in the moving $B$-meson frame with $n_z \cdot v \gg 1$.
To this end, it will be instructive to understand the characteristic  feature of
$\varphi_B^{+}(\xi, \mu)$ with distinct non-perturbative models of $\phi_{B}^{+}(\omega, \mu)$.
Taking advantage of the two phenomenological models motivated by the HQET sum rule calculation
at leading order (LO) \cite{Grozin:1996pq} and at next-to-leading order (NLO) \cite{Braun:2003wx}
\begin{eqnarray}
\phi_{B, \, {\rm I}}^+(\omega, \, \mu=1.5 \, \rm{GeV}) &=& \frac{\omega}{\omega_0^2}e^{-\omega / \omega_0}, \nonumber \\
\phi_{B, \, {\rm II}}^+(\omega, \, \mu=1.5 \, \rm{GeV}) &=&
\bigg [ \frac{1}{k^2+1} - \frac{2 \, (\sigma_B^{(1)} -1)}{\pi^2} \, \ln k \bigg ] \nonumber \\
&& \times \frac{4}{\pi \, \omega_0} \, \frac{k}{k^2+1}, \,\,
k = \frac{\omega}{1.0 \, \rm{GeV}},  \nonumber \\
\label{two models of B-meson LCDA}
\end{eqnarray}
the obtained QCD factorization formula (\ref{QCDF-formula}) implies the shapes of $\varphi_B^{+}(\xi, \mu)$
displayed in figure \ref{models of quasi-DA} at different values of $n_z \cdot v$ ,
where the reference values of the logarithmic inverse moments $\omega_0 = 350 \, {\rm MeV}$
and $\sigma_B^{(1)} = 1.4$ are taken  for the illustration purpose.
It is evident that $\varphi_B^{+}(\xi, \mu)$ develops a radiative tail at large momentum $\xi$ irrespective of
the functional form of $\phi_{B}^{+}(\omega, \mu)$ and,
in contrast to the quasi-parton distribution function (PDF) \cite{Stewart:2017tvs},
no peaks emerge in the momentum region $\xi \leq 0$.
We also mention in passing that the predicted shapes of the leading-twist
$B$-meson LCDA $\phi_{B}^{+}(\omega, \mu)$ at large $\omega$ from Ji's proposal can already
confront with the perturbative QCD calculations carried out in \cite{Lee:2005gza,Kawamura:2008vq,Feldmann:2014ika}
and it will thus provide  interesting insight into the parton-hadron duality ansatz adopted in
constructing perturbative QCD factorization theorems.

Implementing the lattice QCD computation of the spatial correlation $\varphi_B^{+}(\xi, \mu)$
in practice will necessitate  (a) reformulation of the hard-collinear factorization theorem (\ref{QCDF-formula})
with either the lattice regularization scheme along the lines of \cite{Zhang:2017bzy}
or the regularization-invariant momentum subtraction (RI/MOM) scheme \cite{Martinelli:1994ty}
as already discussed  in the context of the LaMET approach
\cite{Alexandrou:2017huk,Green:2017xeu,Chen:2017mzz,Lin:2017ani,Stewart:2017tvs};
(b) improvement of various systematic uncertainties generated by the finite lattice spacing
and the finite lattice box as well as  by truncating the Fourier transformation from  coordinate space
with evaluations for a finite number of discrete $\tau$'s to momentum space.
In addition, computing the yet higher-order perturbative correction to the short-distance
Wilson coefficient $H$ and constructing the subleading-power factorization formula for
the equal-time correlation function $\varphi_B^{+}(\xi, \mu)$ will be also in high demand
for precision determinations of the small-momentum behaviours of the $B$-meson
LCDA $\phi_{B}^{+}(\omega, \mu)$.

It would  be also of interest to construct  a complementary method of
determining the $B$-meson distribution amplitude  $\phi_B^{+}(\omega, \mu)$
from the numerical simulation of the following Euclidean quantity in QCD
\begin{eqnarray}
&& i f_B \, m_B \, \psi_{B}^{+}(x, \mu)
=  \int_{- \infty}^{+ \infty} \, { d \tau \over 2 \, \pi }\,
e^{i \, n_z \cdot p \, x  \,  \tau} \,  \nonumber \\
&& \langle  0 | \left (\bar q \, W_n \right)(\tau \, n_z) \,  \not \! n_z \, \gamma_5 \,
\left (W_n^{\dag} \, b \right )(0)| \bar B(p)\rangle \,.
\label{def: twist-2 quasi-B-meson DA in QCD}
\end{eqnarray}
In this respect, the mostly non-perturbative renormalization method
to renormalize the QCD heavy-light current presented in
\cite{ElKhadra:1997hq,Hashimoto:1999yp,ElKhadra:2001rv,Harada:2001fi}
can be applied for the practical calculations,
while in HQET a concept for the non-perturbative
renormalization scheme in position space has been 
presented in  \cite{Korcyl:2015xmd}.
From the perspectives of the continuum QCD, the newly introduced distribution amplitude $\psi_{B}^{+}(x, \mu)$
can be further matched onto  the  Euclidean  HQET quantity  $\varphi_B^{+}(\xi, \mu)$
by integrating out the  short-distance fluctuations  at the heavy-quark mass scale,
in analogy to the hard-collinear factorization formula obtained in  \cite{Ishaq:2019dst,Zhao:2019elu}.
Furthermore, determining hadronic distribution functions on the light-shell can be also achieved by constructing  the
spatial correlation functions of suitable local partonic currents and by establishing the desired QCD factorization
formulae in coordinate space directly \cite{Braun:2007wv,Bali:2017gfr,Bali:2018spj,Detmold:2005gg,Detmold:2018kwu}.

In comparison with the previous computation of $\phi_B^{+}(\omega, \mu)$ from the QCDSR method \cite{Braun:2003wx},
the inapplicability of achieving a direct determination of the distribution amplitude locally in momentum space
due to the singular behaviours of the non-perturbative quark-gluon condensate contributions
has been resolved by combining the lattice simulation of the quasi-distribution amplitude $\varphi_B^{+}(\xi, \mu)$
and the established hard-collinear factorization formula (\ref{QCDF-formula}).
Since our major objective is to explore the opportunity of
accessing the light-cone dynamics of the leading-twist $B$-meson distribution amplitude
$\phi_B^{+}(\omega, \mu)$ starting from the Euclidean space
instead of carrying out a dedicated lattice calculation of the proposed
quasi-distribution amplitude $\varphi_B^{+}(\xi, \mu)$, it prevents us  from drawing a definite
conclusion on the theory precision of the obtained shape of   $\phi_B^{+}(\omega, \mu)$.
Actually, the numerical simulations of such quasi-distribution amplitudes are still at an exploratory stage,
even for the ones suitable for the determination of the light-meson DA.
However, it might not be implausible to expect that  the achieved accuracy of the obtained
$B$-meson DA with our prescription can be comparable to the collinear pion DA
(for the time being,  the estimated errors of the shape parameters $a_2^{\pi}$
and $\delta_2^{\pi}$ are at the level of (30-50)\% as concluded in \cite{Bali:2018spj})  in the long run,
provided that the desired methodologies to control both the statistical errors and the systematic uncertainties
(see \cite{Cichy:2018mum,Alexandrou:2019lfo,Ji:2020ect} for an elaborate discussion on lattice challenges)
can be eventually constructed with the further development of new algorithms and  computing techniques on the lattice
(see, for instance  \cite{Zhang:2019qiq}).

In particular, the fairly encouraging results from the state-of-art computations
of  the nucleon PDFs and the light-meson distribution amplitudes
as summarized in the comprehensive review \cite{Ji:2020ect} evidently demonstrate
that the newly constructed  LaMET formalism allows for
a promising future to systematically compute a wide range of ``parton observables"
with the  demanding computational  resources and the  tremendous development
of new techniques  and algorithms, which enable us to address the different sources of
systematic uncertainties   step in step in the future.
Consequently, the obtained hard-collinear factorization formula for the $B$-meson quasi-distribution
amplitude in HQET will pave the way for the first-principle determination of the desperately desired
LCDA $\phi_B^{+}(\omega, \mu)$, which is undoubtedly of the highest importance to improve
the present theory precision for predicting any exclusive $B$-meson decay observable
standing out constantly as the central focus of the ongoing experimental programs.
In addition, inspecting the nontrivial relations between the two-particle and three-particle $B$-meson
distribution amplitudes due to the QCD equations of motion \cite{Kawamura:2001jm,Kawamura:2001bp,Braun:2017liq}
and the improved OPE constraints for $\phi_B^{+}(\omega, \mu)$ \cite{Kawamura:2008vq}
with the updated values of $\lambda_{E}$ and $\lambda_{H}$ \cite{Nishikawa:2011qk}
by taking advantage of the forthcoming results obtained from the current strategy
can be of substantial significance to boost our confidence on the robustness
of the lattice HQET technique
(see \cite{Hashimoto:1995in,Sloan:1997fc,Foley:2002qv,Horgan:2009ti} for an alternative formalism).

{\bf Conclusion:}
To summarize, we have proposed a novel approach to determine the momentum dependence
of the leading-twist  $B$-meson LCDA $\phi_{B}^{+}(\omega, \mu)$ in bHQET without introducing
any approximation or assumption for its functional form.
Applying the auxiliary heavy-quark field formalism, we have demonstrated  explicitly the
multiplicative renormalizability of the quasi-distribution function $\varphi_B^{+}(\xi, \mu)$
at all orders in QCD. The perturbative matching function entering the hard-collinear
factorization formula of the spatial correlation was further extracted
with the OPE technique at ${\cal O}(\alpha_s)$.
The present strategy of constructing the light-cone distribution functions
in effective field theories can be also applied  to  the various $B$-meson shape functions
relevant to the QCD description of $\bar B \to X_{d, s} \, \gamma$ and to the heavy-baryon distribution
amplitudes appearing in the soft-collinear effective theory (SCET) computation
of $\Lambda_b \to \Lambda \, \ell \, \ell$. Our results are apparently of importance for exploring
the  delicate flavour structure of the SM and beyond at the LHCb and Belle II experiments.

\vspace{0.3 cm}

\begin{acknowledgments}

We are grateful to Vladimir Braun, Tomomi Ishikawa,  Xiang-Dong Ji, Yi-Zhuang Liu,  Yu-Sheng Liu,
Anatoly Radyushkin and Iain Stewart
for illuminating discussions, and to Vladimir Braun and Anatoly Radyushkin for valuable comments on the manuscript.
W.W and J.X. are supported in part by  the National Natural Science Foundation of China
under Grant No. 11575110,  11735010, and 11911530088,  the Natural Science Foundation of Shanghai under Grant  No. 15DZ2272100,
and the Key Laboratory for Particle Physics, Astrophysics and Cosmology, Ministry of Education  of China.
Y.M.W acknowledges support from the National Youth Thousand Talents Program,
the Youth Hundred Academic Leaders Program of Nankai University, the National Natural Science Foundation of China with
Grant No. 11675082 and 11735010, and the Natural Science Foundation of Tianjin with Grant No. 19JCJQJC61100.
The work  of S.Z is supported by Jefferson Science Associates, LLC under
U.S. DOE Contract \#DE-AC05-06OR23177 and by U.S. DOE Grant \#DE-FG02-97ER41028.
We are also thankful to the Mainz Institute for Theoretical Physics (MITP) of the DFG Cluster of Excellence
PRISMA${}^{+}$ (Project ID 39083149) for its hospitality and support.

\end{acknowledgments}



\begin{thebibliography}{99}



\bibitem{Grozin:1996pq}
  A.~G.~Grozin and M.~Neubert,
  Phys.\ Rev.\ D {\bf 55} (1997) 272
  [hep-ph/9607366].



\bibitem{Beneke:1999br}
  M.~Beneke, G.~Buchalla, M.~Neubert and C.~T.~Sachrajda,
  Phys.\ Rev.\ Lett.\  {\bf 83} (1999) 1914
  [hep-ph/9905312].



\bibitem{Beneke:2000wa}
  M.~Beneke and T.~Feldmann,
  Nucl.\ Phys.\ B {\bf 592} (2001) 3
  [hep-ph/0008255].





\bibitem{Beneke:2001at}
  M.~Beneke, T.~Feldmann and D.~Seidel,
  Nucl.\ Phys.\ B {\bf 612} (2001) 25
  [hep-ph/0106067].



\bibitem{Bosch:2001gv}
  S.~W.~Bosch and G.~Buchalla,
  Nucl.\ Phys.\ B {\bf 621} (2002) 459
  [hep-ph/0106081].




\bibitem{Becher:2005fg}
  T.~Becher, R.~J.~Hill and M.~Neubert,
  Phys.\ Rev.\ D {\bf 72} (2005) 094017
  [hep-ph/0503263].




\bibitem{Khodjamirian:2006st}
  A.~Khodjamirian, T.~Mannel and N.~Offen,
  Phys.\ Rev.\ D {\bf 75} (2007) 054013
  [hep-ph/0611193].




\bibitem{DeFazio:2005dx}
  F.~De Fazio, T.~Feldmann and T.~Hurth,
  Nucl.\ Phys.\ B {\bf 733} (2006) 1;
   Erratum: [Nucl.\ Phys.\ B {\bf 800} (2008) 405]
  [hep-ph/0504088].



\bibitem{Wang:2015vgv}
  Y.~M.~Wang and Y.~L.~Shen,
  Nucl.\ Phys.\ B {\bf 898} (2015) 563
  [arXiv:1506.00667 [hep-ph]].



\bibitem{Wang:2017jow}
  Y.~M.~Wang, Y.~B.~Wei, Y.~L.~Shen and C.~D.~L\"{u},
  JHEP {\bf 1706} (2017) 062
  [arXiv:1701.06810 [hep-ph]].




\bibitem{Lu:2018cfc}
  C.~D.~L\"{u}, Y.~L.~Shen, Y.~M.~Wang and Y.~B.~Wei,
  JHEP {\bf 1901} (2019) 024
  [arXiv:1810.00819 [hep-ph]].





\bibitem{Gao:2019lta}
  J.~Gao, C.~D.~L\"{u}, Y.~L.~Shen, Y.~M.~Wang and Y.~B.~Wei,
  arXiv:1907.11092 [hep-ph].




\bibitem{Lange:2003ff}
B.~O.~Lange and M.~Neubert,
Phys.\ Rev.\ Lett.\  {\bf 91} (2003) 102001
[hep-ph/0303082].





\bibitem{Bell:2013tfa}
G.~Bell, T.~Feldmann, Y.~M.~Wang and M.~W.~Y.~Yip,
JHEP {\bf 1311} (2013)  191
[arXiv:1308.6114 [hep-ph]].


\bibitem{Braun:2014owa}
V.~M.~Braun and A.~N.~Manashov,
Phys.\ Lett.\ B {\bf 731} (2014)  316
[arXiv:1402.5822 [hep-ph]].



\bibitem{Braun:2019wyx}
  V.~M.~Braun, Y.~Ji and A.~N.~Manashov,
  Phys.\ Rev.\ D {\bf 100} (2019)  014023
  [arXiv:1905.04498 [hep-ph]].






\bibitem{Lee:2005gza}
S.~J.~Lee and M.~Neubert,
Phys.\ Rev.\ D {\bf 72} (2005)  094028
[hep-ph/0509350].



\bibitem{Kawamura:2008vq}
  H.~Kawamura and K.~Tanaka,
  Phys.\ Lett.\ B {\bf 673} (2009) 201
  [arXiv:0810.5628 [hep-ph]].



\bibitem{Feldmann:2014ika}
T.~Feldmann, B.~O.~Lange and Y.~M.~Wang,
Phys.\ Rev.\ D {\bf 89} (2014) 114001
[arXiv:1404.1343 [hep-ph]].



\bibitem{Braun:2003wx}
V.~M.~Braun, D.~Y.~Ivanov and G.~P.~Korchemsky,
Phys.\ Rev.\ D {\bf 69} (2004)  034014
[hep-ph/0309330].





\bibitem{Beneke:2011nf}
M.~Beneke and J.~Rohrwild,
Eur.\ Phys.\ J.\ C {\bf 71} (2011)  1818
[arXiv:1110.3228 [hep-ph]].


\bibitem{Wang:2016qii}
Y.~M.~Wang,
JHEP {\bf 1609} (2016)  159
[arXiv:1606.03080 [hep-ph]].


\bibitem{Wang:2018wfj}
Y.~M.~Wang and Y.~L.~Shen,
JHEP {\bf 1805} (2018) 184
[arXiv:1803.06667 [hep-ph]].


\bibitem{Beneke:2018wjp}
M.~Beneke, V.~M.~Braun, Y.~Ji and Y.~B.~Wei,
JHEP {\bf 1807} (2018)  154
[arXiv:1804.04962 [hep-ph]].




\bibitem{Ji:2013dva}
X.~Ji,
Phys.\ Rev.\ Lett.\  {\bf 110} (2013)  262002
[arXiv:1305.1539 [hep-ph]].




\bibitem{Ji:2014gla}
X.~Ji,
Sci.\ China Phys.\ Mech.\ Astron.\  {\bf 57}  (2014)  1407
[arXiv:1404.6680 [hep-ph]].






\bibitem{Cichy:2018mum}
  K.~Cichy and M.~Constantinou,
  Adv.\ High Energy Phys.\  {\bf 2019} (2019) 3036904
  [arXiv:1811.07248 [hep-lat]].






\bibitem{Zhao:2018fyu}
Y.~Zhao,
Int.\ J.\ Mod.\ Phys.\ A {\bf 33} (2019) 1830033
[arXiv:1812.07192 [hep-ph]].



\bibitem{Ji:2020ect}
X.~Ji, Y.~Liu, Y.~Liu, J.~Zhang and Y.~Zhao,
[arXiv:2004.03543 [hep-ph]].


\bibitem{Beneke:2005gs}
  M.~Beneke and D.~S.~Yang,
  Nucl.\ Phys.\ B {\bf 736} (2006) 34
  [hep-ph/0508250].




\bibitem{Kawamura:2018gqz}
  H.~Kawamura and K.~Tanaka,
  PoS RADCOR {\bf 2017} (2018) 076.





\bibitem{Fleming:2007qr}
  S.~Fleming, A.~H.~Hoang, S.~Mantry and I.~W.~Stewart,
  Phys.\ Rev.\ D {\bf 77} (2008) 074010
  [hep-ph/0703207].




\bibitem{Gervais:1979fv}
  J.~L.~Gervais and A.~Neveu,
  Nucl.\ Phys.\ B {\bf 163} (1980) 189.



\bibitem{Dotsenko:1979wb}
  V.~S.~Dotsenko and S.~N.~Vergeles,
  Nucl.\ Phys.\ B {\bf 169} (1980) 527.




\bibitem{Falk:1992fm}
  A.~F.~Falk, M.~Neubert and M.~E.~Luke,
  Nucl.\ Phys.\ B {\bf 388} (1992) 363
  [hep-ph/9204229].



\bibitem{Maiani:1991az}
  L.~Maiani, G.~Martinelli and C.~T.~Sachrajda,
  Nucl.\ Phys.\ B {\bf 368} (1992) 281.



\bibitem{Beneke:1994sw}
  M.~Beneke and V.~M.~Braun,
  Nucl.\ Phys.\ B {\bf 426} (1994) 301
  [hep-ph/9402364].



\bibitem{Radyushkin:2017cyf}
  A.~V.~Radyushkin,
  Phys.\ Rev.\ D {\bf 96} (2017)   034025
  [arXiv:1705.01488 [hep-ph]].





\bibitem{Orginos:2017kos}
  K.~Orginos, A.~Radyushkin, J.~Karpie and S.~Zafeiropoulos,
  Phys.\ Rev.\ D {\bf 96} (2017)   094503
  [arXiv:1706.05373 [hep-ph]].




\bibitem{Braun:2018brg}
  V.~M.~Braun, A.~Vladimirov and J.~H.~Zhang,
  Phys.\ Rev.\ D {\bf 99} (2019)   014013
  [arXiv:1810.00048 [hep-ph]].





\bibitem{Craigie:1980qs}
  N.~S.~Craigie and H.~Dorn,
  Nucl.\ Phys.\ B {\bf 185} (1981) 204.




\bibitem{Ji:2017oey}
  X.~Ji, J.~H.~Zhang and Y.~Zhao,
  Phys.\ Rev.\ Lett.\  {\bf 120} (2018)   112001
  [arXiv:1706.08962 [hep-ph]].



\bibitem{Manohar:2000dt}
  A.~Manohar and M.~Wise,
  ``{\it Heavy quark physics},''
  Camb.\ Monogr.\ Part.\ Phys.\ Nucl.\ Phys.\ Cosmol.\  {\bf 10} (2000) 1.



\bibitem{Zhang:2018diq}
  J.~H.~Zhang, X.~Ji, A.~Sch\"{a}fer, W.~Wang and S.~Zhao,
  Phys.\ Rev.\ Lett.\  {\bf 122} (2019)  142001
  [arXiv:1808.10824 [hep-ph]].



\bibitem{Polyakov:1980ca}
  A.~M.~Polyakov,
  Nucl.\ Phys.\ B {\bf 164} (1980) 171.





\bibitem{Stewart:2017tvs}
  I.~W.~Stewart and Y.~Zhao,
  Phys.\ Rev.\ D {\bf 97} (2018)  054512
  [arXiv:1709.04933 [hep-ph]].




\bibitem{Zhang:2017bzy}
  J.~H.~Zhang, J.~W.~Chen, X.~Ji, L.~Jin and H.~W.~Lin,
  Phys.\ Rev.\ D {\bf 95} (2017)   094514
  [arXiv:1702.00008 [hep-lat]].



\bibitem{Martinelli:1994ty}
  G.~Martinelli, C.~Pittori, C.~T.~Sachrajda, M.~Testa and A.~Vladikas,
  Nucl.\ Phys.\ B {\bf 445} (1995) 81
  [hep-lat/9411010].




\bibitem{Alexandrou:2017huk}
  C.~Alexandrou, K.~Cichy, M.~Constantinou, K.~Hadjiyiannakou, K.~Jansen, H.~Panagopoulos and F.~Steffens,
  Nucl.\ Phys.\ B {\bf 923} (2017) 394
  [arXiv:1706.00265 [hep-lat]].




\bibitem{Green:2017xeu}
  J.~Green, K.~Jansen and F.~Steffens,
  Phys.\ Rev.\ Lett.\  {\bf 121} (2018)  022004
  [arXiv:1707.07152 [hep-lat]].



\bibitem{Chen:2017mzz}
  J.~W.~Chen, T.~Ishikawa, L.~Jin, H.~W.~Lin, Y.~B.~Yang, J.~H.~Zhang and Y.~Zhao,
  Phys.\ Rev.\ D {\bf 97} (2018)  014505
  [arXiv:1706.01295 [hep-lat]].


\bibitem{Lin:2017ani}
  H.~W.~Lin {\it et al.} [LP3 Collaboration],
  Phys.\ Rev.\ D {\bf 98} (2018)  054504
  [arXiv:1708.05301 [hep-lat]].



\bibitem{ElKhadra:1997hq}
  A.~X.~El-Khadra, A.~S.~Kronfeld, P.~B.~Mackenzie, S.~M.~Ryan and J.~N.~Simone,
  Phys.\ Rev.\ D {\bf 58} (1998) 014506
  [hep-ph/9711426].



\bibitem{Hashimoto:1999yp}
  S.~Hashimoto, A.~X.~El-Khadra, A.~S.~Kronfeld, P.~B.~Mackenzie, S.~M.~Ryan and J.~N.~Simone,
  Phys.\ Rev.\ D {\bf 61} (1999) 014502
  [hep-ph/9906376].



\bibitem{ElKhadra:2001rv}
  A.~X.~El-Khadra, A.~S.~Kronfeld, P.~B.~Mackenzie, S.~M.~Ryan and J.~N.~Simone,
  Phys.\ Rev.\ D {\bf 64} (2001) 014502
  [hep-ph/0101023].





\bibitem{Harada:2001fi}
  J.~Harada, S.~Hashimoto, K.~I.~Ishikawa, A.~S.~Kronfeld, T.~Onogi and N.~Yamada,
  Phys.\ Rev.\ D {\bf 65} (2002) 094513;
   Erratum: [Phys.\ Rev.\ D {\bf 71} (2005) 019903]
  [hep-lat/0112044].



\bibitem{Korcyl:2015xmd}
  P.~Korcyl, C.~Lehner and T.~Ishikawa,
  PoS LATTICE {\bf 2015} (2016) 254
  [arXiv:1512.00069 [hep-lat]].



\bibitem{Ishaq:2019dst}
  S.~Ishaq, Y.~Jia, X.~Xiong and D.~S.~Yang,
  arXiv:1905.06930 [hep-ph].



\bibitem{Zhao:2019elu}
  S.~Zhao,
  Phys.\ Rev.\ D {\bf 101} (2020) 7,  071503
  [arXiv:1910.03470 [hep-ph]].




\bibitem{Bali:2018spj}
  G.~S.~Bali {\it et al.},
  Phys.\ Rev.\ D {\bf 98} (2018)  094507
  [arXiv:1807.06671 [hep-lat]].





\bibitem{Braun:2007wv}
  V.~Braun and D.~M\"{u}ller,
  Eur.\ Phys.\ J.\ C {\bf 55} (2008) 349
  [arXiv:0709.1348 [hep-ph]].




\bibitem{Bali:2017gfr}
  G.~S.~Bali {\it et al.},
  Eur.\ Phys.\ J.\ C {\bf 78} (2018)  217
  [arXiv:1709.04325 [hep-lat]].






\bibitem{Detmold:2005gg}
  W.~Detmold and C.~J.~D.~Lin,
  Phys.\ Rev.\ D {\bf 73} (2006) 014501
  [hep-lat/0507007].




\bibitem{Detmold:2018kwu}
  W.~Detmold, I.~Kanamori, C.~J.~D.~Lin, S.~Mondal and Y.~Zhao,
  PoS LATTICE {\bf 2018} (2018) 106
  [arXiv:1810.12194 [hep-lat]].






\bibitem{Alexandrou:2019lfo}
C.~Alexandrou, K.~Cichy, M.~Constantinou, K.~Hadjiyiannakou, K.~Jansen, A.~Scapellato and F.~Steffens,
Phys.\ Rev.\ D {\bf 998} (2019)  114504
[arXiv:1902.00587 [hep-lat]].






\bibitem{Zhang:2019qiq}
  R.~Zhang, Z.~Fan, R.~Li, H.~W.~Lin and B.~Yoon,
  Phys.\ Rev.\ D {\bf 101} (2020)   034516
  [arXiv:1909.10990 [hep-lat]].




\bibitem{Kawamura:2001jm}
  H.~Kawamura, J.~Kodaira, C.~F.~Qiao and K.~Tanaka,
  Phys.\ Lett.\ B {\bf 523} (2001) 111;
  Erratum: [Phys.\ Lett.\ B {\bf 536} (2002) 344]
  [hep-ph/0109181].


\bibitem{Kawamura:2001bp}
H.~Kawamura, J.~Kodaira, C.~Qiao and K.~Tanaka,
Int.\ J.\ Mod.\ Phys.\ A  {\bf 18}  (2003) 1433
[arXiv:hep-ph/0112146 [hep-ph]].





\bibitem{Braun:2017liq}
  V.~M.~Braun, Y.~Ji and A.~N.~Manashov,
  JHEP {\bf 1705} (2017) 022
  [arXiv:1703.02446 [hep-ph]].


\bibitem{Nishikawa:2011qk}
T.~Nishikawa and K.~Tanaka,
Nucl.\ Phys.\ B {\bf 879}  (2014) 110
[arXiv:1109.6786 [hep-ph]].



\bibitem{Hashimoto:1995in}
S.~Hashimoto and H.~Matsufuru,
Phys.\ Rev.\ D {\bf 54} (1996) 4578
[arXiv:hep-lat/9511027 [hep-lat]].






\bibitem{Sloan:1997fc}
J.~H.~Sloan,
Nucl.\ Phys.\ B Proc.\ Suppl.\  {\bf 63} (1998) 365
[arXiv:hep-lat/9710061 [hep-lat]].





\bibitem{Foley:2002qv}
K.~M.~Foley and G.~P.~Lepage,
Nucl.\ Phys.\ B Proc.\ Suppl.\   {\bf 119} (2003) 635
[arXiv:hep-lat/0209135 [hep-lat]].



\bibitem{Horgan:2009ti}
R.~Horgan, L.~Khomskii, S.~Meinel, M.~Wingate, K.~Foley, G.~Lepage,
V.~Hippel, G.M., A.~Hart, E.~Muller, C.~Davies, A.~Dougall and K.~Wong,
Phys.\ Rev.\ D {\bf 80} (2009) 074505
[arXiv:0906.0945 [hep-lat]].




\end{thebibliography}
\end{document}